# Ising model: Elementary excitation and nonsingular heat capacity at critical temperature

You-gang Feng

*Department of Basic Sciences, College of Science, Guizhou University, Huaxi , Guiyang 550025, China*

We find a new parameter vector $q$ to describe spin correlations and fluctuation characteristics. The conservation of scalar $q$ indicates there are simple harmonic motions of $q$, and the motion quantum is called block-spin phonon like the phonon in a crystal, resulting in nonsingular heat capacity at $T_c$. The harmonic motions show there are hierarchies and symmetries of fluctuations, and the soft modes lead to the interactions of block-spin phonons of different frequencies.

## 1. Introduction

The critical phenomena are characterized by the free energy singularity and the long-range correlations of spins. The former indicates that an old phase, a disordered state, disappears; the latter shows that this is just the property of a new phase, an order state. As a new state it is considered a normal ferromagnetic that should have its normal and nonsingular free energy and heat capacity. It is incredible that a ferromagnetic Ising model has no normal thermodynamic quantity to that spins contribute. So far there has been only one theory, the spin-wave theory, can explain these normal quantities.

 Bloch firs postulated the theory in the 1930's [1]. He assumed that the spin wave consisted of a single reversed spin distributed coherently over a large number of otherwise aligned atomic spins in a one-dimensional lattice system. He found that the low-energy excited states of a ferromagnetic would be of this character. The resulting thermodynamic properties of a ferromagnetic at low temperature were in agreement with experiment. Dyson argued that Bloch neglected the spin wave interactions and there was no ferromagnetism in a one-dimensional chain of spins. He revised the Bloch's theory and drew a physical picture for a *3*-dimensional model [2]. As his showing, on microscopic level spin waves arise from interference effects in the lowest partial-wave collisions resulting in a rotation of the spins of the scattered atoms. The

E-mail: ygfeng45@yahoo.com.cn

cumulative effect of many such-rotating collisions is such that inhomogeneous spin states propagate like waves rather than diffusively. Dyson presented that his theory was suitable to those when the temperature was far below the critical temperature. Vaks and his colleagues tried to apply the theory to a Heisenberg model to study the spin waves and correlation function [3]. They found that the damping of the spin waves and the damping increased with the increasing temperature in the range $T < T_c$. All of these tell us that the spin-wave picture cannot illustrate the critical phenomena, especially the critical properties of Ising model, since there is no spin-spin collision caused by the moving atoms with spins.

In this paper we try to show by our theory the normal properties of the new phase for Ising models at $T_c$. In section 2 we find a spin parameter vector $\boldsymbol{q}$ describing the block-spin correlations and get a conservation equation of scalar $q$, revealing the block-spin correlations exist in the form of simple harmonic waves. As a result we get the quantum of the wave motion, the block-spin phonon. In section 3 we obtain nonsingular heat capacities for the new phase and discuss the correlation functions with some symmetric properties and interpret the hierarchies and soft modes in the fluctuations. Section 4 is conclusion.

## 2. Theory

### 2.1 Spin parameter vector $q$

It has become clear to us that a system will never arrive at its critical point due to that the self-similar transformation forbids the fractal side $n^*$ [4]. Even if the critical point relies on an integer side in the hexagonal lattice system those lattice spins outside blocks result in the fluctuation in side around the critical point. Now that we are able to have a system being at $T_c$, that the singularity of free energy at $T_c$ having been proven to be a fact by experimental observation comes from the correlations of the block spins with integer side $n$ near $n^*$ may be a rational explanation. When the original lattice system becomes order the transformations require that there are only at every moment those block spins with the same magnitude $S$ as concerned in a certain side $n$. Consider two limit cases. In the first, a finite number $r$, $0 < r < +\infty$, on the $r$-th hierarchy the vector summation of block spins is always zero although the spins are correlative with each other. In the meantime a system appears order only on the infinite hierarchy, namely $r \to +\infty$, where the system is just an isolated block spin after infinite iteration of the transformations. Denote the system magnetization by $\boldsymbol{M}_1$. In the second case, on a finite hierarchy all block spins are parallel to each other, consequently the system is ordered. Denote the relevant magnetization by $\boldsymbol{M}_2$. Obviously, the magnitude $M_1$ is smaller than the magnitude $M_2$. So the $M_1$ is connected with $T_c$, and the $M_2$ with a temperature $T$ lower than $T_c$.

The fluctuations not only are the deviation in the block sides, but also are the deviation in the block-spin states. An obvious disadvantage of the conventional spin parameter $S$ is that it is always one-dimension such that it cannot be a match for the spin correlations that have the same dimensions as a $d$-dimensional lattice system. In order to research for the fluctuation nature we should find a new spin parameter of $d$ dimensions, besides $S$, to describe the correlations.

Solving the Gaussian model we introduce a parameter $q$ in the Fourier transform. A new lattice spin is expressed by $S_i = (1/\Omega)\sum_q S_q \exp q r_i$ or $S_q = V \sum_i S_i \exp{-q r_i}$, see equation (11) of [4]. The vector $q$ is a reciprocal lattice vector for the new lattice but the one for a block spin system. The magnitude $q$ of $q$ determines the new lattice spin magnitude being consistent with the magnitude of a certain block spin, its changes in both direction and magnitude relate to the changes of the new lattice spins in both direction and magnitude. Because there is a mapping relationship between a new lattice spin and a block spin, if the vector $q$ can serve as an appropriate parameter to demonstrate the block spin state for a certain block-spin system instead of the new lattice spin system, it should have specific features that there is an one-to-one correspondence between the length $q$ of $q$ and the magnitude $S$ of a block spin, each block spin has its own $q$, and all of block spins with $q$ are correlated. The vector's change in direction is connected with the block spin change in direction. The traversal time of the component of $q$, $q_x$, or $q_y$, or $q_z$, in either its own positive-direction state or negative-direction state is identical, since every block spin has the same probability in both the spin-up state and the spin-down state in the thermodynamic equilibrium on any finite hierarchy to keep the spin vector summation to be zero. In a word, the $q$ will be a $d$-dimensional periodically varying parameter rather than random.

**2.2 Conservation equation of scalar $q$**

On the one hand the free energy becomes singularity at $T_c$; on the other hand the fluctuations go on around the critical point, such that the magnitude of the block spin is not the minimum related to the critical point, and $q$ doesn't vanish at $T_c$, $q = 0$ linking to $K_c$. Therefore, the algebraic expression the equation (19) in the equation (21) of [4] changes into

$$K(q) = 1/<S_{tr}^2> \quad , \quad <S_{tr}^2> = S_{tr}^2 \neq S_{tr,\min}^2 \tag{1}$$

Where $K(q) = K_c \sum_\delta \exp{-q \, \delta_{ij}}$, $K_c = 1/2D_{tr,\min}^2$. The $q$ changes around the point

$q = 0$. For the trigonal lattice, we make series expansion for $K(\boldsymbol{q})$ about $q = 0$, and keep the constant and the quadratic terms of $q$, and get

$$q_x^2 + q_y^2 = 4[1 - 1/(6K_c S_{tr}^2)]/a^2 \tag{2}$$

For the trigonal lattice block spin, $a = n + 1$, $6S_{tr}^2 = 1/2D_{tr}$, see the equation (7) of [4].

We then get a conservation equation of scalar $q$ for every block spin of $n$ and $D_{tr}$:

$$q_x^2 + q_y^2 = 4(1 - D_{tr,\min}/D_{tr})/(n+1)^2 \tag{3}$$

Where $q = 2(1 - D_{tr,\min}/D_{tr})^{1/2}/(n+1)$ is constant for a certain block. The equation shows that there is a circle of radius $q$ and a rotating vector $\boldsymbol{q}$, and the initial point and the terminal point of vector are at the circle center and on the circumference, respectively. When $\boldsymbol{q}$ rotates its direction changes and the rotation is a unique motion way for $\boldsymbol{q}$. It is important to point out that the vector $\boldsymbol{q}$ in equation (3) is no longer a reciprocal lattice vector of the Gaussian model, for the magnitude of $\boldsymbol{q}$ is not conservative in the model. Equation (3) fits into each block spin of a block spin system at $T_c$. Clearly, the following set of the simple harmonic vibrations is the simplest solution of equation (3):

$$q_x = q\cos\omega t \quad , \quad q_y = q\sin\omega t \tag{4}$$

Where $\omega$ is a circular frequency, $t$ is time. Now that the spins are correlative with one another but random, the introduction of time parameter is an inevitable result. The properties of trigonometric functions guarantee $\boldsymbol{q}$ to be a good parameter. Generally, a block spin can be written as the form

$$S(\boldsymbol{q},\boldsymbol{r}) = S_q \exp q_x x \exp q_y y \tag{5}$$

as a term in the Fourier transform [4], where $S_q = S$ is the magnitude of a block spin with fractal dimensions $D_{tr}$ in the original spin orientation. Equation (5) states that when the block spins are correlative with each other the spin states propagate in the waves along each dimensional direction. $S(\boldsymbol{q},\boldsymbol{r})$ doesn't change the original spin orientation, its state bears on the exponential functions in equation (5). On substitution of equation (4) in equation (5), $S(\boldsymbol{q},\boldsymbol{r})$ becomes a compound function. Its form, however, is too complicated and redundant to serve as an explicitly visual description of the spin correlations.

For the simplicity, we can directly select $\boldsymbol{q}$ as a pure parameter to describe the spin correlations. A set of simple harmonic waves also plays role of solution of equation (3):

$$q_x = q\cos[\omega(t \mp r/v) + \alpha] \quad , \quad q_y = q\sin[\omega(t \mp r/v) + \alpha] \tag{6}$$

Where the sign "-" or "+" represents the forward wave or the backward wave, the system can be in either the one state or the another state; $v$ is the wave velocity (the system is a uniform medium), $r = (x^2 + y^2)^{1/2}$, $x$ and $y$ the position coordinates of the symmetric center of a block, $\alpha$ the initial phase angle of the block at site $r = 0$. The initial condition is

$$\cos^2(\mp\omega r/v + \alpha) + \sin^2(\mp\omega r/v + \alpha) = 1 \tag{7}$$

In order to illustrate explicitly the correspondence between $q$ and the block spin state, consider a particular situation: Let the spin direction be parallel to the $y$-axis, and $q_y > 0$ refers to the spin-up state, $q_y < 0$ to the spin-down state. A block spin travels in each state for the same time, half period. The traversal time of a spin at the state $q_y = 0$ is omitted. Let the component $q_y$ be independent, and $q_x$ follow after it by equation (6). Clearly, without $q_x$ there is no harmonic motion of $q$ although $q_x$ is not of independence. Let us investigate the correlation of two block spins; they may be either adjacent or far apart; and denote their parameters by $q_1$ and $q_2$, the vectors rotate anticlockwise. At the moment $t_1$, their components in the $y$-axis are positive, $q_{1y} > 0$, $q_{2y} > 0$, the spins are parallel up; at $t_2 (t_2 > t_1)$, $q_{y1} < 0$ and $q_{y2} > 0$, anti-parallel; at $t_3 (t_3 > t_2)$, $q_{1y} < 0$ and $q_{2y} < 0$, parallel down; at $t_4 (t_4 > t_3)$, $q_{1y} > 0$ and $q_{2y} < 0$, anti-parallel, the spin states are just opposite to their states at $t_2$; at $t_5 (t_5 > t_4)$, $q_{1y} > 0$ and $q_{2y} > 0$, parallel up, but $q_1$ and $q_2$ are not the same as those at $t_1$; at $t_6 (t_6 > t_5)$, both vectors return to the states at $t_1$, indicating the two block spins come back their states at $t_1$. The time difference $\Delta t = t_6 - t_1$ is just a vibration period.

It is easy to prove that the locus of $q$ in a 3-dimensional block spin system is a sphere of radius $q$. We can in particular think of the spin-up state as $q_z > 0$, the spin-down state as $q_z < 0$. A set of trigonometric functions will depict the behaviors of rotating vector $q$.

## 2.3 Block-spin phonon

Note that there is no parameter describing a wave state in between two nearest neighbors in equation (6) that means we regard in fact the system as continuum. It is well known that a simple harmonic wave is an elastic wave. The elastic wave in the continuum has the specific properties of acoustic wave satisfying the long-wavelength limit [5].

In the crystal the same harmonic motion of lattices can be respectively described by several waveforms for different purposes. Similarly, let us consider the harmonic motion of $q$ from another angle of view. For the cube ordered reducible block spin system denote the position coordinates of the symmetric center of a block by $(x, y, z)$, every block has its own center. All of these centers make up a $3$-dimensional lattice system with lattice constant $a = n+1$, where $n$ is the block side. For the elastic vibration of the $q$ at site $(x_p, y_p, z_p)$ as being the position of the $p$-th block, there is an effective elastic force $f_p$, a restoring force that may be driven by the fluctuation-dissipation mechanism. Because the Hook's law the force $f(x_p)$, the component of $f_p$ in the x-axis, caused by the displacements the adjacent $q_x(x_{p+1}, t)$ and $q_x(x_{p-1}, t)$ relative to the $q_x(x_p, t)$, respectively, is given by

$$f(x_p) = C[q_x(x_{p+1}, t) - q_x(x_p, t) + q_x(x_{p-1}, t) - q_x(x_p, t)]$$
$$= M \, \mathrm{d}^2 q_x(x_p, t) / \mathrm{d}t^2 \tag{8}$$

Where $C$ is a proportional constant, $M$ effective mass, all displacements have the time dependence $\exp -\omega t$. Since the harmonic motion leads to a dynamic equation $\mathrm{d}^2 q_x(x_p, t)/\mathrm{d}t^2 = -\omega^2 q_x(x_p, t)$, using that equation (8) then becomes a difference equation in the displacements of $q_x$ and has traveling wave solution of the

$$q_x(x_{p\pm 1}, t) = q \exp pka \exp \pm ka \tag{9}$$

Where $k$ is the magnitude of wave-vector $\mathbf{k}$, $x_p = pa$. Combining equation (8) with (9), and using $\mathrm{d}^2 q_x(x_p, t)/\mathrm{d}t^2 = -\omega^2 q_x(x_p, t)$, we get $\omega^2 = (2C/M)(1 - \cos ka)$. As the long-wavelength limit $ka \ll 1$, we expand $\cos ka \cong 1 - (ka)^2/2$, and get a dispersion relation linking up the frequency $\omega$ and $k$

$$\omega^2 = (C/M)k^2 a^2 \tag{10}$$

Where $v = \omega/k$, $v$ is the acoustic velocity. So does the motion in the y-axis or in the z-axis. The previous procedure is completely analogous to the dealing with lattice wave in a crystal [5], so that it is easy to make quantization for the wave motion of *q*, and the quantum of the motion is called the block-spin phonon like the phonon in the lattice wave. For the brevity, we don't write here the process.

For the cube lattice system, the interaction between nearest-neighbor sub-blocks is along the directions normal to the sub-block side (long side), see figure 3 of [4]. Infinite sub-blocks construct a *2*-dimensional system because the symmetric centers of these sub-blocks are on the same plane. There are many such planes parallel to one another, on each of them there is a sub-block spin system, and there is no any interaction between the systems since the directions of sub-block interactions are parallel to the planes. It is easy to prove that in every such system there are also the simple harmonic waves of *q* similar to as shown by equation (6), and there are also sub-block spin phonons in each system like the block-spin phonons in the ordered reducible block-spin system. For simplicity, we call the sub-block spin phonon the block-spin phonon.

The excitation phonon number reaches the maximum at $T_c$, the paired number of anti-parallel spins does the maximum too, such that the block spin vector summation and the one of sub-block spins vanish on any finite hierarchy. Clearly, both the numbers decrease as temperature decreases, leading to that the spin vector summation is no longer zero on a finite hierarchy. As a result, the system magnetization strengthens.

## 3. Discussion

### 3.1 Nonsingular heat capacities

Consider a sub-block spin system of the cube lattice: the symmetric centers of the sub-blocks form a *2*-dimensional lattice system of area *G*. The vibration number of *q* per dimensionality from *k* to $k+\mathrm{d}k$ is given by $2G\pi k\,\mathrm{d}k$. It can equivalently be, using equation (10) and identity $v = \omega/k$, expressed by $2\pi Gk\,\mathrm{d}k = 2\pi G(\omega/v^2)\,\mathrm{d}\omega$ for $\omega \to \omega + \mathrm{d}\omega$. The phonons obey the Plank distribution [5]. We then get the total average phonon energy $<E>$ for a sub-block system

$$<E> = (4\pi G/v^2)k_B T(k_B T/\hbar)^2 \int_0^{x_D}[x^2/(e^x - 1)]\,\mathrm{d}x \qquad (11)$$

Where $\hbar$ is Plank constant, $x = \hbar\omega/(k_B T)$, $x_D = \hbar\omega_D/(k_B T)$, $\omega_D$ is Debye frequency. Note that the total number of the sub-block systems equals $V_s/G(n+1)$, where $V_s$ is the original lattice system volume, the space of adjacent planes is $(n+1)$, *n* the sub-block side. The heat capacity of all sub-block systems is

$$C_v = [V_s/G(n+1)](\partial <E>/\partial T)_G$$

$$= [4\pi V_s / v^2(n+1)]k_B(k_B T)^2 \int_0^{x_D} [x^2/(e^x - 1)]\,dx \tag{12}$$

We expand temporarily the temperature to $T$ slightly below $T_c$ when make calculus, and the integral value is finite. Equation (12) states that $C_v$ obey the $T^2$ law and can be shared by all of *2*-dimensional systems including the trigonal lattice system. With the same reason, for the ordered reducible block spin system of cube lattice the heat capacity behaves as the $T^3$ law:

$$C_v = (12\pi V_s / v^3)k_B(k_B T)^3 \int_0^{x_D} [x^4 e^x/(e^x - 1)^2]\,dx \tag{13}$$

Where the integral value is finite. The law of equation (13) is suitable to all of *3*-dimensional systems. The cube lattice system includes two types of interactions involving the sub-block spins and the ordered block spins, respectively, such that the heat capacity takes the following form of

$$C_v = AT^2 + BT^3 \tag{14}$$

Where the *A* and the *B* are positive constants.

Differing from the block-spin phonon model the heat capacities of the lattice waves observe the $T^3$ law or the $T^2$ law (for the planar systems) only at low temperature that may be far away from $T_c$, since only there the long-wavelength limit holds [5].

### 3.2 Symmetries and hierarchies of fluctuations

There are a few of papers to study the correlation functions at $T_c$ for Ising models. G. Delfino and G. Mussardo studied the spin-spin correlation function in the *2*-dimensional Ising model in a magnetic field at $T_c$, they used the scattering method applied usually in the spin-wave theory and considered the electron charge action, more or less apart from the phase transition topic [6]. Crag A. Tracy and Barry M. McCoy presented the spin-spin correlation by considering the neutron scattering effect and making use of phenomenological formula, yet the fluctuation fine structure was a riddle to us [7]. The problem is that if we have not a good spin parameter to describe the fluctuation nature, we will not be able to find the function with specific form to portray the critical behavior, and the function is very the correlation one. In terms of the Prigogine's theory of self-organization [8], the new phase at $T_c$ in essence is the self-organization in the thermodynamic equilibrium as a sequence of the space-time order of spins. It is well known that there is conservation there are some symmetries. The conservation of $q$ reveals the space-time symmetric properties of the fluctuations at $T_c$. In the following we think of the trigonal lattice as an example for the discussion. The trigonometric functions in equation (6) may be equal to a kind of spin correlation functions at $T_c$.

Although every block spin changes its state constantly, all of the changes show a certain harmonization. The terminal point of $\boldsymbol{q}$ circles about the center $q=0$ obeying the symmetric properties of a rotating group and leading directly to the space-time order of spin states: The time translation invariance: a certain spin state will periodically occur at the same position. The space translation invariance: the same state will simultaneously turn up at some sites regularly arranged. Though different sides correspond to different circles their radii depend on the sides, these centric circles have the same symmetries but chaos. In addition, by the scaling law so long as the hierarchy number $r$ is finite the space of adjacent blocks or adjacent sub-blocks is always the lattice constant, such that the fluctuations will take place on any hierarchies. Thus, the fluctuations are of hierarchies. In this sense, we may say that the fluctuations itself are the very new phase characteristics.

### 3.3 Soft modes

Soft modes exist in vibration systems, originating from local crystal defects, lattice deformation and anharmonic forces, such that the mode phenomena can occur in a wide variety of problems such as lattice wave, ferromagnetic, ferroelectric, superconductor, quantum phase transition, DNA structure, etc. [9-16], and therefore the fluctuations of Ising models at $T_c$ will be comprised in that.

When we Fourier transformed $K(\boldsymbol{q})$ about $q=0$ we neglected the term of $q^4$ that is next to the term of $q^2$ as the function cosine is even. Counting the term of $q^4$, an effective potential energy is given by

$$U(q_i) = Cq_i^2 - fq_i^4 \tag{15}$$

Where $i = x, y, z$, $C$ and $f$ are positive constants, the term of $q_i^2$ leads to harmonic force linked to harmonic vibration, and the term of $q_i^4$ to soft vibration. Like in the lattice wave where the anharmonic term is concerned in thermal expansion [17-18], the term of $q_i^4$ results in nonzero average value of $q$, $<q>\neq 0$, on the finite hierarchy, but $<q>=0$ for the harmonic motion. The result marks that as the anharmonicity the lattice spin system can become order on any finite hierarchy rather than on infinite hierarchy. In addition, when the system adjusts the block side in order to approach to the critical point further new blocks will come out, and old blocks will be decomposed as well, as appropriate for the modulation of the lattice constant spacing adjacent block symmetric centers. Deductively, there exist interactions of block-spin phonons with different frequencies, like the phonon interactions in a crystal [17-18]. We note that in the lattice wave model because the soft mode effect is too weak to affect the existence of the phonon model and the Debey heat capacity that has been in agreement with experiment. The reason may be suitable to the block-spin phonons. In a word, soft modes are of the fluctuation characteristic.

## 4. Conclusion

The vector *q* is a good spin parameter; it shows that the spin correlations at $T_c$ behave as simple harmonic motion, and the motion quantum results in nonsingular heat capacity. The fluctuations are of symmetries and hierarchies, being just the characteristics of the order phase. There are soft modes in the fluctuations, leading to the interactions between block-spin phonons of different frequencies.